\documentclass[12pt] {article} \usepackage{graphicx} \usepackage{epsfig}
\begin {document} \parindent=15pt
\begin{center} {\bf OPEN CHARM PRODUCTION IN $pp$ AND HEAVY ION COLLISIONS
IN QCD}\\ \vspace{.5cm} C. Merino, C. Pajares, and Yu. M. Shabelski$^*$
\\ \vspace{.5cm} Departamento de F\'\i sica de Part\'\i culas, Facultade
de Fisica, and Instituto Gallego de Altas Energias (IGAE), Universidade
de Santiago de Compostela, \\ E-15706-Santiago de Compostela, Spain \\
\vspace{.5cm} $^*$ Permanent address: Petersburg Nuclear Physics Institute,
\\ Gatchina, St.Petersburg 188350 Russia \\ \end{center} \vspace{.5cm}

\begin{abstract}

The RHIC data on charm production are compared with the $k_T$-factorization 
approach  predictions, both standard NLO QCD and FONLL. The calculated
results underestimate the STAR Collaboration data. The role of possible
nuclear effects is discussed.  

\end{abstract}

\vskip .3cm

The investigation of heavy quark production in high energy collisions is an
important method for studing the quark-gluon structure of hadrons and the
possible nuclear effects at early stages of secondary production. The 
description of hard interactions in hadron collisions within the framework of 
QCD is possible only with the help of some phenomenological assumptions which 
reduce the hadron--hadron interaction to the parton--parton one via the 
formalism of the hadron structure functions. The cross sections of hard 
processes in hadron--hadron interactions can be written as the convolutions of 
squared matrix elements of the subprocess calculated within the framework of 
QCD with the parton distributions in the colliding hadrons.

The most popular phenomenological approach is the NLO QCD collinear 
approximation [1-4], where the 
cross sections of QCD subprocesses are calculated in the Next-to-Leading Order 
(NLO) of $\alpha_s$ series. The Fixed Order plus Next-to-Leading-Log (FONLL) 
\cite{CGN} also resumes large perturbative terms proportional to 
$\alpha_s^n \log^k(p_T/m)$ with $k = n, n-1$, where $m$ is the heavy quark 
mass. In these calculations all particles involved are assumed to be on mass 
shell, carrying only longitudinal momenta, and the cross section is averaged 
over two transverse polarizations of the incident gluons. The virtualities 
$q^2$ of the initial partons are taken into account only through their 
structure functions.

The standard QCD expression for heavy quark production cross section in a 
hadron~1 - hadron~2 collision has the form 
\begin{equation}
\sigma^{12\rightarrow Q\overline{Q}} = \int_{x_{a0}}^{1} dx_a
\int_{x_{b0}}^{1} dx_b \cdot G_{a/1}(x_a,\mu^2_F) \cdot G_{b/2}(x_b,\mu^2_F) 
\cdot \hat{\sigma}^{ab\rightarrow Q\overline{Q}}(\hat{s},m_Q,\mu^{2}_R) \;,
\end{equation}
where $\mu_F$ is the QCD factorization scale, 
$x_{a0} = \textstyle 4m_Q^2/ \textstyle s$, and 
$x_{b0} = \textstyle 4m_Q^2/ \textstyle (sx_a)$. Here $G_{a/1}(x_a,\mu^2_F)$ 
and $G_{b/2}(x_b,\mu^2_F)$ are the structure functions of partons $a$ and $b$ 
inside hadrons $1$ and $2$ respectively, and 
\begin{equation}
\hat{\sigma}^{ab\rightarrow Q\overline{Q}}(\hat{s},m_Q,\mu^{2}_R) =
\alpha^2_s(\mu^2_R) \, \sigma_{ab}^{(0)} + \alpha^3_s(\mu^2_R) \, 
\sigma_{ab}^{(1)}
\end{equation}
is the cross section of the subprocess $ab\rightarrow Q\overline{Q}$ as given 
by standard QCD as a sum of LO and NLO contributions \cite{NDE,Alt,Bee}. 
These contributions depend on the parton center-of-mass energy 
\mbox{$\hat{s} = (p_a+p_b)^2 = x_ax_bs$}, the mass of the produced heavy quark 
$m_Q$ (actually they only depend on $\rho = 4m^{2}_{Q}/\hat{s}$), and the QCD 
renormalization scale $\mu^2_R$.  

The possibility to incorporate the incident parton transverse momenta is 
refered to as the $k_T$-factorization approach [6--9], or the theory of 
semihard interactions [10--18]. Here the Feynman diagrams are calculated 
taking into account the virtualities and all possible polarizations of the 
incident partons. In the small $x$ domain there are no ground to neglect the 
transverse momenta of gluons, $q_{1T}$ and $q_{2T}$, when compared to the 
quark mass and transverse momenta, $p_{iT}$. Moreover, at very high energies 
and very high $p_{iT}$ the main contribution to the cross sections comes from 
the region of $q_{1T} \sim p_{1T}$ or of $q_{2T} \sim p_{1T}$ (see  \cite{RSS} 
for details). The QCD matrix elements of the partonic subprocesses are rather 
complicated in such an approach. We have 
calculated them in LO. On the other hand, the multiple emission of soft gluons 
is included here. All details of the calculations in the $k_T$-factorization 
approach, presented below, can be found in \cite{SS}. In our calculations the
charm quark mass was taken as $m_c =$ 1.4 GeV and we used QCD scales
$\mu^2_R = \mu^2_F = m^2_T$, $m^2_T = m^2_c + p^2_T$. 
 
We will firstly consider the cross sections of charm production in $pp$ 
collisions and then we will briefly discuss the situation with nuclear effects.

The existing data on total cross section of charm production at high energies
are presented in Fig.~1. One can see the difference in results by PHENIX and 
by STAR Collaborations. This difference is discussed in details in 
\cite{Xu,Sua}.

\begin{figure}[htb]
\centering
\includegraphics[width=.55\hsize]{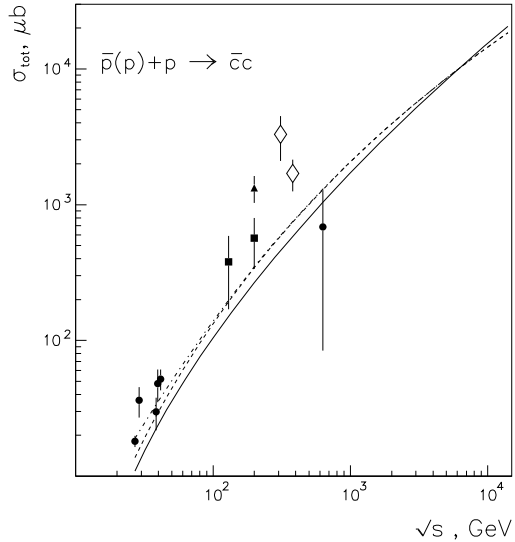}
\vskip -.5cm
\caption{
Total cross section of charm production in $pp$ and $\bar{p}p$ collisions at 
$p \geq 400$ GeV/c. Experimental fixed target data are  taken from \cite{Lou} 
and \cite{Abr}, UA2 Collaboration data from \cite{UA2}, all they are shown by 
points. Squares correspond to PHENIX data [22, 23] and triangle to STAR data 
[24]. The data of cosmic rays taken from  \cite{Xu} are shown by diamonds.
The solid curve shows the result of $k_T$-factorization approach, dashed curve 
corresponds to NLO QCD with only gluon-gluon fusion contribution, and 
dash-dotted curve to total NLO QCD result.}
\end{figure}

All experimental points except those by STAR Collaboration and the cosmic ray 
ones are in reasonable agreement with NLO QCD calculations (dash-dotted
curve), where the GRV95 parton distributions \cite{GRV} were used, and which 
are compatible with more modern analysis (see discussion in \cite{GRV1}). The 
$k_T$-factorization approach result (solid curve) underestimates the data at 
comparatively low energies because it contains only gluon-gluon fusion 
contribution and is close to the NLO QCD collinear approximation curve at 
higher energies.

STAR Collaboration obtains also the $p_T$-distributions of $D$-mesons 
produced in d+Au collisions and scaled to $pp$ collisions at $\sqrt{s}$ = 
200 GeV. These data, taken from \cite{CNV} are presented in Fig.~2. together
with theoretical calculations. The NLO QCD result shown by dashed curve
is in evident disagreement with the data on the same level as in Fig.~1. The 
upper curve of FONLL calculations for charmed quark production taken from 
\cite{CNV} (dash-dotted curve) also underestimates the data. The result of the
$k_T$-factorization approach (solid curve) has reasonable slope in 
$p_T$-dependence but also underestimates the data. The level of disagreement 
in this case is, however, smaller than in the case of total cross sections. 
This should be connected with the fact that a rather large contribution to the 
total cross section comes from low-$p_T$ region ($p_T \leq 1$ GeV/c) where 
both experimental and theoretical uncertainties are rather large.

\begin{figure}[htb]
\centering
\includegraphics[width=.55\hsize]{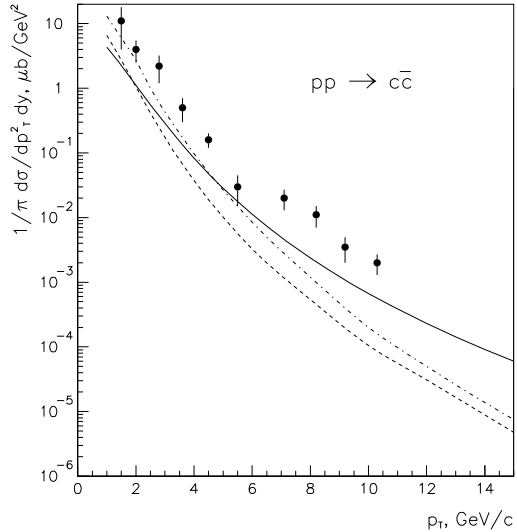}
\vskip -.5cm
\caption{
STAR Collaboration data for $p_T$-distributions of $D$-mesons produced in
d + Au collisions and scaled to $pp$ collisions at $\sqrt{s}$ = 200 GeV
together with calculations in $k_T$-factorization approach (solid curve), NLO 
QCD (dashed curve), and FONLL \cite{CNV} (dash-dotted curve).}
\end{figure}

It is necessary to note that in Fig.~2 we compare the experimental points for 
$D$-meson production with theoretical curves for charmed quark distributions. 
Contrary to the mention in \cite{CNV}, together with fragmentation processes 
where the momentum of $D$-meson is smaller than the momentum of $c$-quark, 
there exist recombination processes where the momentum of $D$-meson is larger 
than the momentum of $c$-quark. The existence of recombination processes in 
charm production seems to be evident from the experimental data on the 
asymmetry in yields of the so-called favoured and unfavoured $D$-mesons, see 
discussions in [30--33]. As a matter of fact the produced heavy and light 
quarks have very different transverse momenta but the difference in the 
components of their velocities can be not so large. Possibly, the 
fragmentation and recombination processes in charm quark hadronization balance 
each other in the processes with not very high $p_T$, e.g. the calculated 
Feynman-$x$ distributions of produced charm quarks in $\pi p$ collisions are 
in good agreement with the experimental distributions of produced $D$-mesons
(see Fig.~5 in \cite{FMNR}).

The total cross section of charm production was not measured at 
Tevatron-collider energies. However there exist data on $p_T$-distributions of 
$D$-mesons at these energies \cite{Aco}. They are presented in Fig.~3 where it
is shown that they are in good agreement with NLO QCD calculations for charm 
quarks (dashed curve) based on GRV95 parton distributions. The solid curve 
corresponds to the $k_T$-factorization approach and it slightly overestimates 
the data, but the agreement should become better in the future when the 
contribution of charmed antibaryons will be added to the experimental data. 
The results of FONLL calculations \cite{CNV} with fragmentation functions for 
$D$-mesons production presented in \cite{Aco}, slightly underestimate the 
yields of $D$-mesons. Thus all QCD approaches (see also \cite{Ram}), except of 
the extremal ones, e.g. the one presented in \cite{DY}, are in reasonable 
agreement with experimental data at $\sqrt{s}$ = 1.96 TeV. 

\begin{figure}[htb]
\centering
\includegraphics[width=.55\hsize]{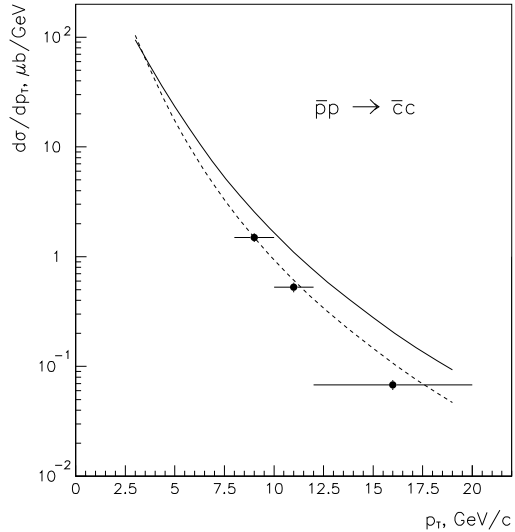}
\vskip -.5cm
\caption{
The cross sections for $D$-meson production in $\bar{p}p$ collisions at 
$\sqrt{s}$ = 1.96 TeV \cite{Aco} with $\vert y_1 \vert \leq 1$ together with 
calculations in $k_T$-factorization approach (solid curve) and NLO QCD (dashed 
curve).}
\end{figure}

Let us shortly discuss the nuclear effects in charm production at RHIC. 
First of all we can say that EMC-effect, i.e. the nuclear deformation of 
parton distributions \cite{ESK} should increase the total cross sections for 
charm production, calculated in the NLO QCD linear approximation at RHIC 
energy by 5-10 \% \cite{APSS} in comparison with a linear A-dependence.

The data of PHENIX and STAR Collaborations scaled to binary interactions are 
presented together in \cite{Sua}. One can see the absence of nuclear effects
from $pp$ to central Au-Au collisions from PHENIX data, and from d-Au to
central Au-Au collisions from STAR data, both on the level of 20\% accuracy.
Concerning the PHENIX data presented in \cite{Xu}, there is no visible
dependence of the charm production cross section on the number of collision
participant nucleons, $N_{part}$, in Au-Au collisions at different 
centralities.

There exists a factor $\sim$ 2-3 discrepancy \cite {Sua} in the total cross 
sections of charm production obtained by PHENIX and STAR Collaborations.
Surely, the explanation of this discrepancy is completely connected to some
experimental problem. However, meanwhile experiments don`t clarify this point,
two possibilities can be imagined: \\
i) if we trust PHENIX data, the NLO QCD reasonably describes all
experimental data except of two cosmic ray points presented in Fig.~1, and then
nuclear effects in total cross sections of charm production are small; \\

ii) if, on the contrary, we trust STAR data, nuclear effects increase the 
total cross sections of charm production about 4-5 times. If the $N_{part}$ 
dependence of these nuclear effects saturates very fast, even the cosmic ray 
data could be included in the theoretical description. The origin of so large 
nuclear effects can be connected with large non-perturbative contributions in 
high density states (which can be larger than perturbative contributions), 
e.g. string fusion \cite{MPR}, percolation \cite{BP}, or colour glass 
condensate effects \cite{GV,KT} in the interactions with nuclei. 

In these approaches, above some scale $\eta_c$ \cite{BP} given by the critical 
percolation string density, the strong colour field inside the cluster formed 
by the overlapping strings produces $Q\bar{Q}$ pairs via the Schwinger 
mechanism as a single string produces light $q\bar{q}$ pairs. In the same way, 
in the colour glass condensate approach \cite{GV,KT} the significant scale is 
the saturation momentum $Q_s$ which grows with energy and nuclear size. When
$Q_s > m_T(c\bar{c})$ , the classical colour field is strong and produce pairs 
$Q\bar{Q}$. The production pattern for heavy quarks becomes similar to that of 
the light quarks and an overall enhancement of heavy quark production cross
section is thus expected. \\

Summary : \\

1. The $k_T$-factorization approach predictions are in reasonable agreement
with NLO QCD for the total cross section of charm production at RHIC energies.
We obtain a reasonable description of the PHENIX data and we are in 
contradiction with STAR data.

2. It seems that the main part of our disagreement with STAR data comes from
low-$p_T$ region ($p_T \leq 1$ GeV/c) where both experimental and
theoretical uncertainties are rather large.

3. The predicted $p_T$-distribution of the produced charm at high $p_T$ in the
$k_T$-factorization approach is higher than NLO QCD and FONLL predictions.
The $k_T$-factorization approach only slightly underestimates the experimental
STAR data.

4. PHENIX data are compatible with the absence of any nuclear effects in 
charm production, whereas STAR data need rather strong nuclear effects at
RHIC energies. \\

We are grateful to N. Armesto, A. Khodjamirian, M. G. Ryskin, and A. G. Shuvaev 
for useful discussions. This work was supported by Ministerio de Educaci\'on y
Ciencia of Spain under project FPA 2005--01963 and by Xunta de Galicia (Conseller\'\i a de
Educaci\'on). It was also supported in part by grant RSGSS--1124.2003.2


\end{document}